\documentclass[11pt]{article}

\newsavebox{\foobox}
\newcommand{\slantbox}[2][0]{\mbox{%
        \sbox{\foobox}{#2}%
        \hskip\wd\foobox
        \pdfsave
        \pdfsetmatrix{1 0 #1 1}%
        \llap{\usebox{\foobox}}%
        \pdfrestore
}}
\newcommand\unslant[2][-.25]{\slantbox[#1]{$#2$}}

\newcommand{\mpi}{\text{\unslant[-.18]\pi}}
\newcommand{\mdelta}{\text{\unslant[-.18]\delta}}

\usepackage[left=2cm, right=2cm, top=2.5cm, bottom=2.5cm]{geometry}
\usepackage[x11names]{xcolor}
\usepackage{fancyhdr, amssymb, cancel, amsmath, graphicx, pgfplots, tikz}
\usepackage{isomath}

\usetikzlibrary{shadows}

\newcommand{\stylecolor}{violet}

\usepackage[labelfont={bf,sf, color=\stylecolor}, margin={1.5cm,0cm}]{caption}

\usepackage[colorlinks=true, urlcolor=\stylecolor!70!white, linkcolor=\stylecolor, citecolor=\stylecolor!70!white, hyperindex=true, linktocpage=true]{hyperref}

\usepackage[explicit]{titlesec}

\newcommand*\sectionlabel{}
\titleformat{\section}
  {\gdef\sectionlabel{}
   \Large\bfseries\scshape}
  {\gdef\sectionlabel{\thesection }}{0pt}
  {\begin{tikzpicture}[remember picture,overlay]
       \end{tikzpicture}
  }
\titlespacing*{\section}{0pt}{0pt}{0pt}

\newcommand*\subsectionlabel{}
\titleformat{\subsection}
  {\gdef\subsectionlabel{}
   \large\bfseries\scshape}
  {\gdef\subsectionlabel{\thesubsection  }}{0pt}
  {\begin{tikzpicture}[remember picture,overlay]
    	\draw (-0.15, 0.02) node[right] {\color{\stylecolor} \textsf{\subsectionlabel}};
	\draw (1.25, 0) node[right] {\color{\stylecolor} \textsf{#1}};
	\fill[color=\stylecolor] (1,-0.25) rectangle (1.1, 0.25);
       \end{tikzpicture}
  }
\titlespacing*{\subsection}{0pt}{10pt}{10pt}

\newcommand*\subsubsectionlabel{}
\titleformat{\subsubsection}
  {\gdef\subsubsectionlabel{}
   \bfseries\scshape}
  {\gdef\subsubsectionlabel{\thesubsubsection.\ \  }}{0pt}
  {\begin{tikzpicture}[remember picture,overlay]
    	\draw (-0.15, 0) node[right] {\color{\stylecolor} \textsf{\subsubsectionlabel#1}};
       \end{tikzpicture}
  }
\titlespacing*{\subsubsection}{0pt}{7pt}{7pt}

\pgfplotsset{every axis legend/.append style={at={(1.02,1)},anchor=north west}}

\begin{document}

\allowdisplaybreaks

\pagestyle{fancy}
\renewcommand{\headrulewidth}{0pt}
\fancyhead{}

\fancyfoot{}
\fancyfoot[C] {\textsf{\textbf{\thepage}}}

\begin{equation*}
\begin{tikzpicture}
\draw (\textwidth, 0) node[text width = \textwidth, right] {\color{white} easter egg};
\end{tikzpicture}
\end{equation*}

\begin{equation*}
\begin{tikzpicture}
\draw (0.5\textwidth, -3) node[text width = \textwidth] {\huge  \textsf{\textbf{Conductivity bounds in  probe brane models}} };
\end{tikzpicture}
\end{equation*}
\begin{equation*}
\begin{tikzpicture}
\draw (0.5\textwidth, 0.1) node[text width=\textwidth] {\large \color{black} \textsf{Tatsuhiko N. Ikeda, Andrew Lucas and Yuichiro Nakai}};
\draw (0.5\textwidth, -0.5) node[text width=\textwidth] {\small\textsf{Department of Physics, Harvard University, Cambridge, MA 02138, USA}};
\end{tikzpicture}
\end{equation*}
\begin{equation*}
\begin{tikzpicture}
\draw (0, -13.1) node[right, text width=0.5\paperwidth] {\texttt{tikeda@physics.harvard.edu \\ lucas@fas.harvard.edu \\ ynakai@physics.harvard.edu}};
\draw (\textwidth, -13.1) node[left] {\textsf{\today}};
\end{tikzpicture}
\end{equation*}
\begin{equation*}
\begin{tikzpicture}
\draw[very thick, color=\stylecolor] (0.0\textwidth, -5.75) -- (0.99\textwidth, -5.75);
\draw (0.12\textwidth, -6.25) node[left] {\color{\stylecolor}  \textsf{\textbf{Abstract:}}};
\draw (0.53\textwidth, -6) node[below, text width=0.8\textwidth, text justified] {\small We discuss upper and lower bounds on the electrical conductivity of finite temperature strongly coupled quantum field theories, holographically dual to probe brane models, within linear response.  In a probe limit where disorder is introduced entirely through an inhomogeneous background charge density, we find simple lower and upper bounds on the electrical conductivity in arbitrary dimensions.  In field theories in two spatial dimensions,  we show that both bounds persist even when disorder is included in the bulk metric.   We discuss the challenges with finding sharp lower bounds on conductivity in three or more spatial dimensions when the metric is inhomogeneous.};
\end{tikzpicture}
\end{equation*}

\tableofcontents

\titleformat{\section}
  {\gdef\sectionlabel{}
   \Large\bfseries\scshape}
  {\gdef\sectionlabel{\thesection }}{0pt}
  {\begin{tikzpicture}[remember picture,overlay]
	\draw (1, 0) node[right] {\color{\stylecolor} \textsf{#1}};
	\fill[color=\stylecolor] (0,-0.35) rectangle (0.7, 0.35);
	\draw (0.35, 0) node {\color{white} \textsf{\sectionlabel}};
       \end{tikzpicture}
  }
\titlespacing*{\section}{0pt}{15pt}{15pt}

\begin{equation*}
\begin{tikzpicture}
\draw[very thick, color=\stylecolor] (0.0\textwidth, -5.75) -- (0.99\textwidth, -5.75);
\end{tikzpicture}
\end{equation*}

\section{Introduction}
One of the simplest experimental probes of strongly interacting quantum phases of matter without quasiparticles is the electrical conductivity $\sigma$.   As $\sigma=\infty$ in a translation-invariant metal (at finite charge density), quantitative theories for $\sigma$ require a careful understanding of the mechanisms of translational symmetry breaking.    One recent trend has been to employ gauge-gravity duality, also called holography \cite{review1, review2, review3},  which allows to compute correlation functions of strongly interacting quantum systems at finite temperature and charge density by mapping them to dual, \emph{classical} computations of perturbing charged black holes. 

In recent years, a predictive theory of transport in such phases has begun to emerge \cite{hkms}.  These ideas have inspired and found applications in holographic \cite{hartnollimpure, hartnollhofman, btv, dsz, lss} and non-holographic \cite{raghu2, patel, patel2} models of strange metals.  In fact, it is now known that when translational symmetry breaking is weak, then holographic, memory function and hydrodynamic approaches give identical results for $\sigma$ \cite{lucas1501, lucasMM, lucas4}.   These models of quantum critical transport in fluids have recently found experimental applications to charge neutral graphene \cite{crossno, lucas3}.

When effects such as disorder cannot be treated perturbatively, one possible outcome is localization, where the electronic wave functions become spatially localized, leading to the vanishing of the conductivity at zero temperature.   The original model of localization was non-interacting electrons hopping on a lattice with random on-site energies.  For electrons hopping on lattices in spatial dimensions $d\le 2$, any amount of randomness causes localization \cite{anderson, abrahams}.   At finite temperature,  the conductivity becomes finite, but is exponentially suppressed \cite{halperin2}.  Recently, it has been pointed out that this effect can survive in an interacting theory \cite{basko} -- this has been coined many-body localization (MBL).   MBL has also attracted much attention from the viewpoint of quantum statistical mechanics \cite{polkovnikov, rigol}, as a counter example to a naive expectation that every interacting quantum system eventually thermalizes.   Furthermore, it has been experimentally realized by using ultracold atoms \cite{bloch1,bloch2}.   Theoretical studies have revealed many aspects of MBL in small systems by numerical exact diagonalization \cite{huse1,huse2,canovi,dalessio,khemani,johri,rigolA,rigolB,agarwal,ponte,gopalakrishnan},  RG \cite{huse3,potter} and entanglement \cite{grover}.   However, almost all that is known about MBL is for disordered spin models in one spatial dimension -- an important open question remains whether MBL is a robust phenomenon in higher spatial dimensions -- and if so, in what models and circumstances.

A natural question to ask is whether holography can fill the gap mentioned above, and give us insight into the possibility for MBL in higher dimensional models.  So far, however, holographic models do not readily predict MBL.   The simplest ``mean field" holographic models of strongly disordered metals predict a diffusion-limited regime with strictly finite transport coefficients \cite{vegh, davison, blake1, donos1, andrade, donos2, gouteraux, gouteraux2, davison15, blake2}.    Inspired by this work, \cite{hartnoll1} proposed that strongly interacting metals may have diffusion-limited transport, in which the role of disorder is relatively minor.\footnote{As noted in \cite{lucas4}, such a diffusion-limited model is, in some respects, similar to resistor network transport models  \cite{kirkpatrick}.}  More recently, it has been shown that broad classes of holographic models (including the ``mean field" models above) admit diffusion-limited transport, in the sense that transport coefficients are bounded from below by universal constants, no matter the nature of disorder \cite{grozdanov, grozdanov2}.    Hence, many of the simplest holographic disordered metals are immune to many-body localization, and  are reminiscent (though not identical) to the proposal of \cite{hartnoll1}.   

In this paper, we will apply these ideas to a different class of holographic models of metals employing probe branes \cite{kobayashi, karch, yaida, hartnoll09, ugajin, sonner}, which we will quickly review in Section \ref{sec2}.   Our main result is that diffusion-limited electrical transport also persists to these probe brane models, which have a more complicated action than the simple model of \cite{grozdanov}.  We discuss how to analytically compute the  conductivity of these models in terms of black hole horizon data in Section \ref{sec3}, along with a discussion of our variational techniques.    The remainder of the paper discusses our conductivity bounds.   We will show that for probe brane models dual to theories in two spatial dimensions, a universal lower bound identical to \cite{grozdanov} holds.    For three (or more) spatial dimensions, simple lower bounds only exist in special limiting cases, which we discuss as well.   We also discuss the possibility for upper bounds on the conductivity.  Unlike in the case of standard Einstein-Maxwell-dilaton holographic models, it is sensible to discuss upper bounds on the conductivity in probe brane models, as we will see below.     

\section{Probe Brane Models}\label{sec2}
Let us now review holographic probe brane models of metals.   These models are based on a Dirac-Born-Infeld (DBI) action for the electromagnetic field, as we review below, and arise naturally in ``top-down" holographic models.  The basic idea is that we will place a small number of charged branes on a charge neutral geometry,  and neglect the backreaction of the latter when discussing the charge dynamics on the probe branes.

Let us begin by reviewing the canonical example of such a probe brane set-up.   The gravity backgrounds upon which we add a small number of probe branes is formed by a large stack of $N_{\mathrm{c}}$ D3 branes within type IIB string theory, in the limit of large 't Hooft coupling $g_{\mathrm{YM}}^2 N_{\mathrm{c}}\equiv \lambda \gg 1$, leads to the well-known near-horizon $\mathrm{AdS}_5\times \mathrm{S}^5$ geometry \cite{maldacena}: \begin{equation}
\mathrm{d}s^2_{10} = L^2\left[\frac{\mathrm{d}r^2}{(r+\mpi T)^2f_T(r)} - f_T(r) (r+\mpi T)^2 \mathrm{d}t^2 + (r+\mpi T)^2\mathrm{d}\mathbf{x}^2\right] + L^2 \mathrm{d}\mathrm{\Omega}_5^2,   \label{eq:Ads5metric}
\end{equation}
with $\mathrm{d\Omega}_5^2$ the unit metric on a round $\mathrm{S}^5$ and \begin{equation}
f_T(r)  = 1-\left(\frac{\mpi T}{\mpi T+r}\right)^4.
\end{equation}
The prefactor $L= \lambda^{1/4}\sqrt{2\mpi \alpha^\prime}$, with $1/2\mpi \alpha^\prime$  the fundamental string tension.    $T$ is the Hawking temperature of the black hole.

Following \cite{karch}, and continuing along with our example of a D3 brane background, we now wrap a small number $N_{\mathrm{f}} \ll N_{\mathrm{c}}$ of D$p$ branes on $\mathrm{AdS}_{d+2}\times \mathrm{S}^{p-1-d}$;  we choose a maximally sized ``equatorial" sphere for simplicity, and assume that the branes do not move along the internal sphere.   As in \cite{karch},  we choose $p=5$ to model metals in $d=2$, and $p=7$ to model metals in $d=3$. The low energy effective action on the D$p$ branes is \begin{equation}
S = -N_{\mathrm{f}}T_{\mathrm{D}p}\int \mathrm{d}^{p+1}x\; \sqrt{X},
\end{equation}with \begin{equation}
X = -\det(g_{ab} + 2\mpi \alpha^\prime F_{ab}).    \label{defX}
\end{equation}
Here $g_{ab}$ is the induced world-volume metric and $F_{ab}$ is the induced world-volume field strength;  the indices $ab$ run over all the worldvolume dimensions on the D$p$ branes.   As $N_{\mathrm{f}} \ll N_{\mathrm{c}}$, we may neglect the backreaction of the D$p$ branes on the geometry.      We may integrate out the directions on the sphere in $S$ and obtain an action dependent only on the $d+2$ AdS dimensions: \begin{equation}
S = -N_{\mathrm{f}}T_{\mathrm{D}p} V_{\mathrm{S}} \int \mathrm{d}^{d+2}x\; \sqrt{X} \equiv -\mathcal{N} \int \mathrm{d}^{d+2}x\; \sqrt{X},  \label{eq:Xact}
\end{equation}
where $V_{\mathrm{S}}$ is the volume of the internal sphere wrapped by the D$p$ branes:  $V_{\mathrm{S}} = L^{p-d-1}\Omega_{p-d-1}$, with $\Omega_{p-d-1}$ the volume of the unit sphere $\mathrm{S}^{p-d-1}$.    The equations of motion associated the DBI action are \begin{equation}
\partial_M \left(\sqrt{X} \left(g + 2\mpi \alpha^\prime F\right)^{MN} - \sqrt{X} \left(g + 2\mpi \alpha^\prime F\right)^{NM}\right)=0, \label{eom}
\end{equation}
with the upper indices in the previous equation used to denote the matrix inverse of $g+2\mpi \alpha^\prime F$.   The $MN\cdots$ indices run over the $d+2$ dimensions of the $\mathrm{AdS}_{d+2}$.   Henceforth, we will allow for $d$ to be general in (\ref{eq:Xact}), as there are other D$q$-D$p$ brane set-ups that are possible for other choices of $d$.

As $N_{\mathrm{f}}\ll N_{\mathrm{c}}$,  fluctuations of $F$ do not backreact on the metric $g$, which we take to be $t$-independent.   On general principles, we now choose to fix the metric to take the form \begin{equation}
\mathrm{d}s^2 = g_{rr}(r,\mathbf{x})\mathrm{d}r^2 + g_{tt}(r,\mathbf{x}) \mathrm{d}t^2 + g_{ij}(r,\mathbf{x})\mathrm{d}x^i \mathrm{d}x^j,  \label{eq:BG1}
\end{equation}
with $i$ and $j$ running over the spatial indices, as well as \begin{equation}
A = A_t(r,\mathbf{x}) \mathrm{d}t.  \label{eq:BG2}
\end{equation}
In particular, this latter ansatz assumes the absence of any worldvolume magnetic flux.   The background (\ref{eq:BG2}) can be found through an exact solution of the DBI equations,  but its precise form will not be necessary in this paper.   

We now pick a convenient coordinate system, where the AdS bulk radial direction is denoted $r$, the black hole horizon is fixed at $r=0$, and an asymptotically AdS region is at $r=\infty$.  Demanding regularity, the near-horizon geometry is \begin{subequations}\label{eq8}\begin{align}
g_{rr} &= \frac{V(\mathbf{x})}{4\mpi Tr} + W_{rr}(\mathbf{x}) + \cdots, \\
g_{tt} &=  - 4\mpi Tr V(\mathbf{x}) - W_{tt}(\mathbf{x})r^2 + \cdots, \\
g_{ij} &= \gamma_{ij} + h_{ij} r + \cdots.
\end{align}\end{subequations} 
$T$ is the Hawking temperature of the black hole, as well as the temperature of the dual boundary theory.  Near the horizon, the background gauge field is \begin{equation}
A_t = \frac{r\beta(\mathbf{x})}{2\mpi \alpha^\prime} + \cdots.
\end{equation}
As we will see below,  $\beta(\mathbf{x})$ is related to the charge density on the black hole horizon.  Near the asymptotically AdS boundary ($r\rightarrow \infty$), the metric becomes \begin{equation}
\mathrm{d}s^2 \approx L^2 \left[ \frac{\mathrm{d}r^2}{r^2} -r^2 \mathrm{d}t^2 + r^2 \mathrm{d}\mathbf{x}^2\right],
\end{equation}
if we assume that the dual field theory lives on flat Minkowski space;  we will not need an explicit expression for the near-boundary behavior of the background gauge field.   Finally, we will assume that the spatial dimensions $\mathbf{x}$ are periodic, forming a $d$-dimensional torus.   

\section{Conductivity} \label{sec3}

Our main goal in this paper will be to compute the direct current electrical conductivity $\sigma$ of the boundary theory, holographically dual to our probe brane set-up.   Namely, if we apply a spatially uniform electric field $\mathbf{E} \mathrm{e}^{-\mathrm{i}\omega t}$ to an isotropic metal,  then \begin{equation}
\langle \mathbf{J}\rangle = \sigma(\omega) \mathbf{E} + \mathrm{O}(\mathbf{E}^3)  \label{ohm}
\end{equation}
with $\mathbf{J}$ the spatial components of a U(1) conserved current, and $\langle \cdots\rangle$ denoting averages over quantum and thermal fluctuations.   This is nothing more than Ohm's Law.  We interpret $\sigma$ as being defined \emph{after} spatially averaging the electric current across the sample.    

On general principles \cite{hkms, lucasMM, davison1507}, the low frequency conductivity of a perfectly clean (translation invariant) metal is \begin{equation}
\sigma(\omega) = \frac{n^2}{\epsilon+P} \left(\mpi \mdelta(\omega) - \frac{1}{\mathrm{i}\omega }\right) + \sigma_{\textsc{q}} + \mathrm{O}(\omega).
\end{equation}
In the above formula, $n$ denotes the thermodynamic charge density,  $\epsilon$ the energy density, $P$ the pressure, and $\sigma_{\textsc{q}}$ a dissipative correction related to charge diffusion.  However, probe brane models do not realize this $\mdelta$ function \cite{karch}.   The basic reason for this is simple; we use the case of the D3-D7 system (dual to metal in $d=3$ spatial dimensions) as an example.  Using the generic relation that $T_{\mathrm{D7}} \sim g_{\mathrm{s}}^{-1} (\alpha^{\prime})^{-4}$ \cite{becker}, $L\sim \lambda^{1/4}\sqrt{\alpha^\prime}$, and $\lambda \sim g_{\mathrm{s}}N_{\mathrm{c}}$ \cite{maldacena},  one can estimate that\footnote{The scaling of $n$ and $\sigma_{\textsc{q}}$ follows from our discussion around (\ref{eq:mempar2}).}  \begin{subequations}\begin{align}
\sigma_{\textsc{q}} &\sim \mathcal{N}(2\mpi \alpha^\prime)^2 L\sim N_{\mathrm{f}}N_{\mathrm{c}}, \\
\frac{n^2}{\epsilon+P} &\sim \frac{\left(\mathcal{N}(2\mpi \alpha^\prime)^2 L\right)^2}{N_{\mathrm{c}}^2} \sim N_{\mathrm{f}}^2,
\end{align}\end{subequations}
where we have employed that the enthalpy $\epsilon+P  \sim N_{\mathrm{c}}^2$ for the D3 branes \cite{witten}, which dominate the enthalpy.   Hence, we see that in the limit $N_{\mathrm{f}} \ll N_{\mathrm{c}}$, the \emph{only} contribution to $\sigma$ is the finite contribution $\sigma_{\textsc{q}}$.   This is related to dissipative charge diffusion in the dual field theory, and so the focus of this paper will be to understand how these dissipative processes are bounded in these top-down probe brane models of holographic metals.

Our calculation will be at strictly finite temperature $T$.   At $T=0$, new $\mdelta$ functions emerge in $\sigma(\omega)$ at finite charge density \cite{hartnoll09}.   However, the coefficient of this $\mdelta$ function is distinct -- for a discussion, see \cite{gouteraux2}.  As we will work at finite temperature, this $\mdelta$ function is smeared, and we include it in $\sigma_{\textsc{q}}$, as we are only computing the conductivity at $\omega=0$.

\subsection{Membrane Paradigm}

To compute the conductivity, we now perturb the background described in the previous section by a small electric field.   In the probe limit, the gauge field is corrected but the geometry is not, and so in linear response we obtain the near-horizon Taylor expansion of the gauge field: \begin{subequations}\label{eq14}\begin{align}
A_t &= \frac{r\beta(\mathbf{x})}{2\mpi \alpha^\prime} - p(\mathbf{x}) - rq(\mathbf{x}) - \cdots, \\
A_r &= -\frac{p(\mathbf{x}) + rq(\mathbf{x})}{4\mpi Tr} + \cdots, \\
A_i &=  -E_i \left(t+\frac{\log (4\mpi Tr)}{4\mpi T}\right) + a_i(\mathbf{x}) + rb_i(\mathbf{x}) + \cdots.
\end{align}\end{subequations}
The functions $p$, $q$, $a_i$, and $b_i$ are all linearly proportional to $E_i$, because the background solution has only $A_t\ne 0$.   The form of these functions is fixed by regularity in infalling coordinates: see e.g. \cite{donos1507}.    Some residual gauge freedom is left, but it is not necessary to fix.  In linear response, the subleading terms above will drop out of the calculation of $\sigma$.

Even at the fully nonlinear level we note some interesting properties of the DBI equations of motion.  Define \begin{equation}
\mathcal{J}^i \equiv  \frac{2\mpi \alpha^\prime\mathcal{N}}{2}\left( \sqrt{X} \left(g + 2\mpi \alpha^\prime F\right)^{ir} - \sqrt{X} \left(g + 2\mpi \alpha^\prime  F\right)^{ri}\right).
\end{equation}
Upper indices imply the matrix inverse in the above equation.   The prefactor is chosen conveniently, for reasons which become clear after (\ref{eq:mempar2}).  The $r$-component of the equations of motion (\ref{eom}) demands that \begin{equation}
\partial_i \mathcal{J}^i = 0.  \label{diji}
\end{equation}
 Furthermore, the $i$-component of (\ref{eom}) implies that \begin{equation}
\partial_r \mathbb{E}[\mathcal{J}^i] = 0,   \label{eq:mempar}
\end{equation}where we have  defined \begin{equation}
\mathbb{E}[\circ] \equiv \frac{1}{L^d} \int \mathrm{d}^d\mathbf{x} \; \circ,
\end{equation}
with $L^d$ the spatial volume of the boundary theory.   Possible boundary terms in (\ref{eq:mempar}) vanish, as the spatial dimensions are compact.  Note that $\mathbb{E}[\circ]$ is not defined in a coordinate independent way, but we will find this helpful for some practical reasons.

Near the asymptotically AdS boundary, $A(r,\mathbf{x})\sim A_0(\mathbf{x}) + A_1(\mathbf{x}) r^{1-d}+\cdots$.  Hence,  $F$ is subleading to $g$ in $\mathcal{J}^i$ as $r\rightarrow\infty$, and so we may Taylor expand \begin{equation}
\mathcal{J}^i (r\rightarrow \infty) \approx L^{d-2} \mathcal{N}( 2\mpi \alpha^\prime )^2 \times  r^d (-\partial_r A_i).   \label{eq:mempar2}
\end{equation}
Using the standard holographic dictionary, this is the \emph{local} expectation value of the current in the boundary theory -- as $r\rightarrow \infty$.    $\partial_r \mathcal{J}^i \ne 0$ locally, but using (\ref{eq:mempar}) we see that $\mathbb{E}[\mathcal{J}^i]$ is independent of bulk radius $r$, and equal to the spatial average of the expectation value of the current operator in the boundary theory.   We may therefore evaluate this average in the bulk near the black hole horizon.   This is exactly the ``membrane paradigm" \cite{iqbal} which has recently been used to reduce holographic dc transport computations to effective fluid dynamical equations on black hole horizons in a large variety of Einstein-Maxwell-dilaton holographic models with inhomogeneous black holes  \cite{donos1409, donos1506, donos1507, donos1511}.

We now compute $\mathcal{J}^i$ at $r=0$, to linear order in $E_i$.    To do this, we re-write the matrix $g+2\mpi \alpha^\prime F$ in blocks, using $g$ from (\ref{eq8}) and $F$ from (\ref{eq14}), separating the $rt$ directions (denoted with $AB$ indices) from the spatial indices $ij$, defining the matrices $Y$, $W$ and $Z$ in the process:\begin{equation}
(g+2\mpi \alpha^\prime F)_{MN} = \left(\begin{array}{cc} Y_{AB} &\  W_{Aj} \\ -(W^{\mathrm{T}})_{Bi} &\ Z_{ij}   \end{array}\right),
\end{equation}
and note that $W = \mathrm{O}(E)$ or $\mathrm{O}(r)$,  whereas $Y$ and $Z$ are non-trivial at leading order.   We denote $Y=Y_0 + \mathrm{O}(E)$, and $Z=Z_0+\mathrm{O}(E)$.   Firstly, we compute \begin{subequations}\begin{align}
(g+2\mpi \alpha^\prime F)^{Aj} &= - Y^{AB} W_{Bi}(Z + W^{\mathrm{T}}Y^{-1}W)^{ij} \approx -Y^{AB}W_{Bi}Z^{ij}, \\
(g+2\mpi \alpha^\prime F)^{iB} &=  (Z+W^{\mathrm{T}} YW)^{ij}(W^{\mathrm{T}})_{jA}Y^{AB} \approx Z^{ij}(W^{\mathrm{T}})_{jA}Y^{AB}
\end{align}\end{subequations} (recall that upper indices imply matrix inverse).   In the last steps above, we have used that the higher order contributions from $W$ either are nonlinear corrections in $E_i$, or vanish on the horizon.   As we approach the horizon at $r=0$, \begin{subequations}\begin{align}
Y &=  \left(\begin{array}{cc} (4\mpi Tr)^{-1} V &\ \beta \\ -\beta &\ -4\mpi T r V  \end{array}\right)\left(1 + \mathrm{O}(r) + \mathrm{O}(E)\right),  \label{eq:18a} \\
Z_{ij} &= \gamma_{ij} + \mathrm{O}(r) + \mathrm{O}(E).
\end{align}\end{subequations}
We hence find that \begin{subequations}\label{eq20}\begin{align}
(g+2\mpi \alpha^\prime F)^{rj} &\approx  \frac{-1}{V^2-\beta^2}\left(\begin{array}{cc} -4\mpi Tr V &\ -\beta    \end{array}\right) \left(\begin{array}{c} -2\mpi \alpha^\prime (4\mpi Tr)^{-1} (E_i - \partial_i p )\\ -2\mpi \alpha^\prime (E_i- \partial_i p)  \end{array}\right)  \gamma^{ij} + \text{subleading} \notag \\
&= -2\mpi \alpha^\prime \gamma^{ji} \frac{(V+\beta)(E_i - \partial_i p)}{V^2-\beta^2} \\
(g+2\mpi \alpha^\prime F)^{ir} &\approx \frac{\gamma^{ij}}{V^2-\beta^2}\left(\begin{array}{cc} -2\mpi \alpha^\prime (4\mpi Tr)^{-1} (E_j- \partial_j p ) &\  -2\mpi \alpha^\prime (E_j- \partial_j p) \end{array}\right) \left(\begin{array}{c} -4\mpi Tr V \\  \beta    \end{array}\right) + \text{subleading} \notag \\
&= 2\mpi\alpha^\prime \gamma^{ij}(E_j  - \partial_j p) \frac{V-\beta}{V^2-\beta^2}
\end{align}\end{subequations}
where we have used the symmetry of $\gamma$ in the last line.   To leading order,\begin{equation}
X(r=0) = -\det(Y) \det\left(Z + W^{\mathrm{T}} Y^{-1}W\right) = -\det(Y_0)\det(Z_0) + \mathrm{O}(E) \approx (V^2-\beta^2) \gamma .
\end{equation}
Recall the definition of $X$ in (\ref{defX});  $\gamma \equiv \det(\gamma_{ij})$.  We do not need to worry about the $\mathrm{O}(E)$ corrections here, as both components of (\ref{eq20}) are $\mathrm{O}(E)$.   Hence, at $\mathrm{O}(E)$: \begin{equation}
\mathcal{J}^i = \frac{\mathcal{N}(2\mpi \alpha^\prime)^2}{\sqrt{1-(\beta/V)^2}} \sqrt{\gamma}\gamma^{ij}(E_j - \partial_j p) .
\end{equation}
Combining this expression with (\ref{diji}) we obtain an equation which we must solve for $p$: \begin{equation}
\partial_i \left[\frac{\mathcal{N}(2\mpi \alpha^\prime)^2}{\sqrt{1-(\beta/V)^2}} \sqrt{\gamma}\gamma^{ij}(E_j - \partial_j p)\right] = 0.  \label{eqdiff}
\end{equation}   Upon doing so,  recalling the definition of $J^i$ in (\ref{ohm}),  we may readily extract the conductivity matrix via \begin{equation}
\mathbb{E}[\mathcal{J}^i] = \mathbb{E}[J^i]\equiv I^i = \sigma^{ij}E_j.
\end{equation}

\subsection{Variational Methods}
In general, we cannot solve for $p$ analytically.   Nonetheless, we can employ variational methods developed in \cite{lucas4} to obtain non-perturbative insight into the behavior of the conductivity in a strong disorder limit, and in particular into whether or not a many-body localized phase is possible.   This non-perturbative insight is based on the key point that (\ref{eqdiff}) is the diffusion equation in an inhomogeneous fluid, with $p$ playing the role of the chemical potential in this abstract fluid.  In particular we may rewrite (\ref{eqdiff}) as \begin{equation}
\nabla^i\left(D(\mathbf{x}) (E_i - \nabla_ip)\right) = 0,  \label{diffeq}
\end{equation}with $\nabla_i$ the covariant derivative with respect to $\gamma_{ij}$, and an effective diffusion constant\begin{equation}
D \equiv \frac{\mathcal{N}(2\mpi \alpha^\prime)^2}{\sqrt{1-(\beta/V)^2}}
\end{equation}
which is inhomogeneous.   We may now use hydrodynamic insight to constrain the resulting conductivities.

On the true solution to (\ref{eqdiff}), $\overline{\mathcal{J}}^i$, we can compute the ``power dissipated" as a local integral over Joule heating: \begin{equation}
\mathcal{P}[\overline{\mathcal{J}}]  \equiv  \mathbb{E}\left[\frac{\sqrt{1-(\beta/V)^2}}{\mathcal{N}(2\mpi \alpha^\prime)^2}\frac{\gamma_{ij}}{\sqrt{\gamma}} \overline{\mathcal{J}}^i \overline{\mathcal{J}}^j\right]. \label{eq24}
\end{equation}
To see why, we use the fact that $\overline{\mathcal{J}}^i = D(E_i - \nabla_ip)$, and find\begin{align} \label{eq25}
\mathcal{P} &= \mathbb{E}\left[ \frac{\sqrt{1-(\beta/V)^2}}{\mathcal{N}(2\mpi \alpha^\prime)^2}\frac{\gamma_{ij}}{\sqrt{\gamma}}  \left\lbrace \frac{\mathcal{N}(2\mpi \alpha^\prime)^2}{\sqrt{1-(\beta/V)^2}} \sqrt{\gamma}\gamma^{ik}(E_k- \partial_k p)  \right\rbrace  \overline{\mathcal{J}}^j \right] \notag \\ 
&= \mathbb{E}\left[\overline{\mathcal{J}}^j (E_j - \partial_j p)\right] = E_j I^j,
\end{align}where we have integrated by parts in the last step above, and employed (\ref{diji}) along with periodic boundary conditions.   Using the definition of $\sigma^{ij}$, we see that \begin{equation}
\mathcal{P}[\overline{\mathcal{J}}]  =  I^i \left(\sigma^{-1}\right)_{ij}I^j.
\end{equation}
This is analogous to the fact that Ohmic heating in a resistor is equal to $I^2R$.     Hence, we can gain information about $\sigma^{-1}$ by computing $\mathcal{P}[\mathcal{J}]$.

Now, suppose that we do not plug in the true current into $\mathcal{P}$, but instead we guess \begin{equation}
\mathcal{J} = \overline{\mathcal{J}} + \widetilde{\mathcal{J}}.
\end{equation}
As (\ref{eqdiff}) is linear,  we can find a solution $\overline{\mathcal{J}}^i$ to the equations of motion such that $\mathbb{E}[\overline{\mathcal{J}}^i] = I^i$.  Hence, we are free to choose the constraint \begin{equation}
\mathbb{E}\left[\widetilde{\mathcal{J}}^i \right] = 0.  \label{jcons1}
\end{equation}
We will also demand that our trial current is conserved:  \begin{equation}
\partial_i \mathcal{J}^i = 0.   \label{jcons2}
\end{equation}
Note that by definition, $\partial_i \overline{\mathcal{J}}^i=0$, and hence $\partial_i \widetilde{\mathcal{J}}^i = 0$.  
An identical manipulation to (\ref{eq25}) reveals that \begin{align}
\mathcal{P}[\overline{\mathcal{J}} + \widetilde{\mathcal{J}}]  &= \mathcal{P}[ \widetilde{\mathcal{J}}]  + \mathcal{P}[\overline{\mathcal{J}} ] + 2\mathbb{E}\left[ \frac{\sqrt{1-(\beta/V)^2}}{\mathcal{N}(2\mpi \alpha^\prime)^2}\frac{\gamma_{ij}}{\sqrt{\gamma}}  \left\lbrace \frac{\mathcal{N}(2\mpi \alpha^\prime)^2}{\sqrt{1-(\beta/V)^2}} \sqrt{\gamma}\gamma^{ik}(E_k- \partial_k p)  \right\rbrace  \widetilde{\mathcal{J}}^j \right] \notag \\
&= \mathcal{P}[ \widetilde{\mathcal{J}}]  + \mathcal{P}[\overline{\mathcal{J}} ] + 2\mathbb{E}\left[ (E_i - \partial_i p ) \widetilde{\mathcal{J}}^i\right]  = \mathcal{P}[ \widetilde{\mathcal{J}}]  + \mathcal{P}[\overline{\mathcal{J}} ] + 2\mathbb{E}\left[  p \partial_i \widetilde{\mathcal{J}}^i \right] + 2 \mathbb{E}\left[\widetilde{\mathcal{J}}^i \right] E_i.
\end{align}
Employing (\ref{jcons1}) and (\ref{jcons2}) the last two terms vanish.  As $\mathcal{P}[\mathcal{J}]\ge 0$ for any $\mathcal{J}$ (as manifest from the definition), we obtain \begin{equation}
\mathcal{P}[\overline{\mathcal{J}} + \widetilde{\mathcal{J}}]  \ge  \mathcal{P}[\overline{\mathcal{J}} ].
\end{equation}
If we suppose for simplicity that our disordered sample is isotropic in the thermodynamic limit, then we immediately find that \begin{equation}
\mathcal{P}[\mathcal{J}] \ge \frac{I^2}{\sigma},  \label{varbound}
\end{equation}
for any possible choice of $\mathcal{J}$ with the proper average current;  furthermore, this bound is saturated only on the true soluion.   Hence,  we may provide lower bounds on $\sigma$ upon inverting (\ref{varbound}).

We may also derive upper bounds on the conductivity \cite{lucas4} by considering the trial function \begin{equation}
\mathcal{P}^\prime[p] = \mathbb{E}\left[\mathcal{N}(2\mpi \alpha^\prime)^2\frac{\sqrt{\gamma}\gamma^{ij}}{\sqrt{1-\beta^2/V^2}}(E_i - \partial_i p)(E_j-\partial_j p)\right].  \label{eq:uppervar}
\end{equation}
We employ the same arguments as before, defining $p=\overline{p}+\tilde p$,  with $\overline{p}$ the solution to (\ref{eqdiff}), though this time with no constraints on $\tilde p$.  It is straightforward to see that $\mathcal{P}^\prime$ is minimized on the true solution to (\ref{eqdiff}), and also that \begin{equation}
\mathcal{P}^\prime \ge \sigma^{ij}E_i E_j \ge \sigma E^2,
\end{equation}
where we have assumed isotropy in the last step.  Hence, we may find both upper and lower bounds on the conductivities.    We will assume isotropy of $\sigma$ for the remainder of the paper, for simplicity.

\section{Disordered Gauge Field}
Let us begin by studying probe branes on the homogeneous AdS-Schwarzschild background with random worldvolume Maxwell flux.   In the boundary theory, this is dual to random fluctuations in the charge density.   This is a particularly simple limit, as \begin{equation}
\gamma_{ij} = L^2 (\mpi T)^2\mdelta_{ij} ,
\end{equation}
for probe brane models with the $\mathrm{AdS}_5\times \mathrm{S}^5$ background.   The computation extends straightforwardly to probe brane models on other backgrounds, if they are homogeneous.  We may relate $\sqrt{\gamma}$ to the entropy density using the Bekenstein-Hawking formula: \begin{equation}
s = \frac{1}{4G_{\mathrm{N}}}\sqrt{\gamma} = \frac{1}{4G_{\mathrm{N}}}\left(L \mpi T\right)^d, 
\end{equation}
with $G_{\mathrm{N}}$ the effective Netwon's gravitational constant in AdS.    Hence, we find that the variational function $\mathcal{P}$ is \begin{equation}
\mathcal{P}[\mathcal{J}] = \mathbb{E}\left[\frac{\sqrt{1-(\beta/V)^2}}{\mathcal{N}(2\mpi \alpha^\prime)^2} \left(\frac{1}{L\mpi T}\right)^{d-2} \mathcal{J}^2\right],
\end{equation}
with $\mathcal{J}$ indices raised and lowered using the Kronecker $\mdelta$.   A simple choice of trial function is \begin{equation}
\mathcal{J}^i = I^i,  \label{eq:simpletrial}
\end{equation}
which leads to \begin{equation}
\frac{1}{\sigma} \le \mathbb{E}\left[\frac{\sqrt{1-(\beta/V)^2}}{\mathcal{N}(2\mpi \alpha^\prime)^2} \left(\frac{1}{L\mpi T}\right)^{d-2}\right].
\end{equation}
As \begin{equation}
\sqrt{1-\frac{\beta^2}{V^2}} \le 1,  \label{eq:sqrtbound}
\end{equation}
we immediately find that
\begin{equation}
\sigma \ge \left(L\mpi T\right)^{d-2} \mathcal{N}(2\mpi \alpha^\prime)^2.  \label{eq:upperbound4}
\end{equation}
(\ref{eq:upperbound4}) confirms for us that these models exhibit diffusion-limited transport, along the lines of \cite{hartnoll1}.   Furthermore, setting $d=2$,  we recover the bound of \cite{grozdanov}.   This bound also is reminiscent of the proposal of \cite{xhge} for $d>2$.   We will return to the question of whether these bounds are robust to disorder in the metric in the next section.

We may find an upper bound on the conductivity employing the simple trial function $p=0$ in $\mathcal{P}^\prime[p]$: \begin{equation}
\sigma \le  \left(L \mpi T\right)^{d-2} \mathcal{N}(2\mpi \alpha^\prime)^2 \mathbb{E}\left[\frac{1}{\sqrt{1-(\beta/V)^2}}\right].  \label{eq:upperbound}
\end{equation}
To interpret this result more naturally, it is helpful to consider the background equation of motion for the gauge field.    Near the horizon,  the $t$-component of the gauge field's equation of motion reads \begin{equation}
0 = \frac{2\mpi \alpha^\prime\mathcal{N}}{2} \partial_r \left(\sqrt{X}(g+2\mpi \alpha^\prime F)^{tr} -\sqrt{X}(g+2\mpi \alpha^\prime F)^{rt}  \right)  + \mathrm{O}(r).
\end{equation}
This is the generalization of Gauss' Law in the DBI system.   Hence, we can interpret the expression, pointwise on the horizon, as the local charge density on the horizon,  which we denote as $n_{\mathrm{h}}$.  Employing (\ref{eq:18a}) as $r\rightarrow 0$  \begin{equation}
n_{\mathrm{h}} = \frac{\mathcal{N}(2\mpi \alpha^\prime)^2 \sqrt{\gamma} \beta}{\sqrt{1-(\beta/V)^2}}.
\end{equation}
Defining a dimensionless charge density \begin{equation}
\tilde n_{\mathrm{h}} \equiv \frac{n_{\mathrm{h}}}{\mathcal{N}(2\mpi \alpha^\prime)^2 V \sqrt{\gamma}}
\end{equation}
which is roughly the horizon charge per unit of entropy, we obtain (using that $\sqrt{1+x^2} \le  1+|x|$, and $\mathbb{E}[|\mathcal{X}|] \le \sqrt{\mathbb{E}[\mathcal{X}^2]}$ from  Jensen's inequality)\begin{equation}
\sigma \le \left(L\mpi T\right)^{d-2} \mathcal{N}(2\mpi \alpha^\prime)^2 \mathbb{E}\left[ \sqrt{1+\tilde n_{\mathrm{h}}^2}\right] \le  \left(L\mpi T\right)^{d-2} \mathcal{N}(2\mpi \alpha^\prime)^2 \left(1+\sqrt{\mathbb{E}\left[\tilde n_{\mathrm{h}}^2\right]}\right),  \label{eq:upperbound2}
\end{equation}
which tells us that the conductivity is bounded from above by the amount of charge density on the horizon, including the effects of spatial fluctuations.

\section{Disordered Geometry}
Next, we turn to the case where the worldvolume metric on the probe branes also is inhomogeneous disorder.   We split this discussion by dimension.

\subsection{$d=1$}
We begin by studying the case $d=1$.   In this limit, we may exactly compute the conductivity, as was done before in \cite{ugajin}.   In this case, the trial function (\ref{eq:simpletrial}) is exact, since $\partial_x \mathcal{J}=0$ is required, and upon employing periodic boundary conditions we find \begin{equation}
\sigma = \frac{ \mathcal{N}(2\mpi \alpha^\prime)^2 }{\mathbb{E}\left[\sqrt{1-(\beta/V)^2}\sqrt{\gamma}\right]}.
\end{equation}

\subsection{$d=2$}
Next, we turn to the case $d=2$.   This case is non-trivial,  but is still special in that we may make a coordinate change to the conformal gauge, where the induced horizon metric is given by \begin{equation}
\gamma_{ij} = G(\mathbf{x})\mdelta_{ij}.
\end{equation}
In this case, using our variational trial current (\ref{eq:simpletrial}), we obtain \begin{equation}
\frac{I^2}{\sigma} \le \frac{1}{ \mathcal{N}(2\mpi \alpha^\prime)^2 } \mathbb{E}\left[\sqrt{1-\frac{\beta^2}{V^2}} \frac{G\mdelta_{ij}}{G} I^i I^j\right] = \frac{I^2}{ \mathcal{N}(2\mpi \alpha^\prime)^2 }\mathbb{E}\left[\sqrt{1-\frac{\beta^2}{V^2}}\right].
\end{equation}
Hence, employing (\ref{eq:sqrtbound}) we obtain \begin{equation}
\sigma \ge \frac{\mathcal{N}(2\mpi \alpha^\prime)^2}{\mathbb{E}\left[\sqrt{1-(\beta/V)^2}\right]} \ge \mathcal{N}(2\mpi \alpha^\prime)^2.
\end{equation}
Note also that this result is identical to what we found in (\ref{eq:upperbound4}) -- namely, it is geometry independent.

In the special case where $\beta=0$, and the black hole is uncharged,  this computation essentially reduces to that in \cite{grozdanov}.   The variational calculation we have described here is simpler than the one presented in \cite{grozdanov}.

Upon employing conformal gauge, the simple trial function $p=0$ again leads to the upper bound (\ref{eq:upperbound}) on $\sigma$, independent of $G$.   The manipulations to obtain (\ref{eq:upperbound2}) are also valid even when the metric is not homogeneous, following the same logic as we employed for the lower bound above.

\subsection{$d>2$}
%

Let us finally make some brief comments on the situation in $d>2$.   Here, we do not expect sharp bounds to exist on $\sigma$, once the metric can be inhomogeneous.   For simplicity, we will focus on the case of uncharged models, and so $\beta=0$.   It is also worth noting that in this limit, the linearized DBI equations of motion reduce to those of Einstein-Maxwell theory, and so our comments in this section are also relevant for Einstein-Maxwell holographic models.

In \cite{xhge} it was proposed that\begin{equation}
\det(\sigma) \ge \left[\mathcal{N}(2\mpi \alpha^\prime)^2 \right]^d \mathbb{E}\left[ \sqrt{\gamma}\right]^{d-2}.  \label{eq:1512bound2}
\end{equation}
The prefactor in front is simply related to the coefficients in our probe brane model, and would be replaced by the Maxwell coupling constant in the Einstein-Maxwell theory.\footnote{Namely, if one expands out the DBI action to quadratic order, one finds $\int \mathrm{d}^{d+2}x \sqrt{-g}(1 + \mathcal{N}(2\mpi \alpha^\prime)^2 F^2/4)+\cdots$.}   Let us now argue that it is possible to violate (\ref{eq:1512bound2}) in $d>2$.   
For simplicity, let us consider black holes where the induced horizon metric takes the form \begin{equation}
\gamma_{ij} = \mathcal{F}(\mathbf{x}) \mdelta_{ij}
\end{equation}
with $\mathcal{F}(\mathbf{x})$ a smooth function related to the local entropy density.  One can use the fluid-gravity correspondence \cite{minwalla} to construct such a black hole explicitly, if the wavelength associated with the inhomogeneity tends to infinity.  In this case, the metric associated with the boundary theory will not be homogenous.   Regardless, we expect that the general mechanism we point out here will be present even when this restriction is lifted, if we backreact other matter content on an AdS-Schwarzschild black hole.

Let us now employ our upper bound (\ref{eq:uppervar}), using the ansatz $p=0$.  Using Jensen's inequality, we see that for any random function $\mathcal{X}\ge 0$,  $\mathbb{E}[\mathcal{X}]^a \le \mathbb{E}[\mathcal{X}^a]$ for $a\ge 1$.   Choosing $a=d/(d-2)$,
\begin{align}
\det(\sigma^{ij}) \le\left[\mathcal{N}(2\mpi \alpha^\prime)^2 \right]^d \mathbb{E}\left[\mathcal{F}^{(d-2)/2}\right]^d \le\left[\mathcal{N}(2\mpi \alpha^\prime)^2 \right]^d\mathbb{E}\left[\mathcal{F}^{d/2}\right]^{d-2} = \left[\mathcal{N}(2\mpi \alpha^\prime)^2 \right]^d \mathbb{E}\left[\sqrt{\gamma}\right]^{d-2}. \label{lasteq}
\end{align}
The first step follows from the fact that this trial function leads to an isotropic $\sigma^{ij}$.  The right-most term in this string of inequalities is the predicted \emph{lower} bound of (\ref{eq:1512bound2}).    Furthermore, one could imagine cooking up boundary conditions with very long wavelength $\mathcal{F}$, but with very large spatial fluctuations in $\mathcal{F}$.  Such large fluctuations would lead to large prefactors relating the expressions in each inequality in (\ref{lasteq}).  It seems possible in principle to obtain therefore a very small ratio of $\det(\sigma)/\mathbb{E}[\sqrt{\gamma}]^{d-2}$, and so we do not expect any simple non-trivial bounds on $\det(\sigma)$ for $d>2$.

\section{Conclusion}
The main goal of this paper has been the demonstration that top-down holographic probe brane models of metals do not readily undergo many-body localization.   Instead, these models are governed by diffusion-limited transport, much like other holographic models \cite{grozdanov, grozdanov2}.   In the course of our calculation, we have extended the ``membrane paradigm" calculations of direct current transport to holographic probe brane models, mimicking similar computations in other holographic models \cite{ugajin, donos1506, donos1507}. 

The probe limit also makes it possible to study the \emph{nonlinear} conductivity, in which $\sigma$ as defined in (\ref{ohm}) becomes $\mathbf{E}$-dependent \cite{karch}.   In particular,  in $d=2$ one finds that the bound (\ref{eq:upperbound4}) holds at the nonlinear level in a \emph{clean} metal, in the absence of a magnetic field.   Though it is straightforward to extend the computation of Section \ref{sec3} to the nonlinear level,  we do not see an obvious way to generalize (\ref{eq:upperbound4}) in the presence of arbitrary spatial inhomogeneity.  

We have also noted the challenge with finding sharp conductivity bounds in higher dimensional theories.   Our results thus call into question a recently proposed universal conductivity bound in holographic models \cite{xhge}.   Whether or not such higher dimensional conductivity bounds can hold in other circumstances is an interesting open question.

One assumption that we made in our probe brane models was that the probe branes wrap a maximally sized sphere on the $\mathrm{S}^5$.   If the size of this sphere shrinks to zero locally, then the branes ``pinch" and the local conductivity can vanish \cite{ugajin}.   It is not clear whether this is an appropriate holographic analogue of many-body localization.   From the point of view of the horizon fluid, this pinching leads to a local depletion of gapless charged excitations.    This seems more analogous to a Mott insulator in condensed matter physics,  than to a disorder-driven localized insulator.   However, the properties of the horizon fluid may not locally coincide with those of the boundary theory.   Hence, this may be an interesting problem to return to with the new techniques that are discussed in this paper.

Finally,  we expect that our bounds may be tested in non-holographic models of quantum critical points.   These may be realized in lattice models that can be simulated using quantum Monte Carlo \cite{wallin}.  Although it is challenging to compute the direct current conductivity accurately, this would provide a highly non-trivial test of holography and of the possibility for diffusion-limited transport in a condensed matter model without any large  $N$ limits.

\section*{Acknowledgements}
We thank Richard Davison for helpful comments.   AL is supported by the NSF under Grant DMR-1360789 and MURI grant
W911NF-14-1-0003 from ARO. TNI and YN are supported by the JSPS Fellowship for Research Abroad.

\begin{appendix}

  \end{appendix}
\bibliographystyle{unsrt}
\addcontentsline{toc}{section}{References}
\bibliography{disorderbib}

\end{document}